\documentclass[twocolumn,showpacs,amsmath,amssymb,prl,superscriptaddress,floatfix]{revtex4-1}
\usepackage{graphicx}
\usepackage{bm}
\usepackage{amsmath,amsfonts}

\begin{document}

\title{Liquid ground state, gap and excited states of a strongly correlated spin chain}

\author{Igor Lesanovsky}

\affiliation{Midlands Ultracold Atom Research Centre (MUARC), School of Physics and Astronomy, The University of Nottingham, Nottingham, NG7 2RD, United Kingdom}


\date{\today}

\begin{abstract}
We present an exact solution of an experimentally realizable and strongly interacting one-dimensional spin system which is a limiting case of a quantum Ising model with long range interaction in a transverse and longitudinal field. Pronounced quantum fluctuations lead to a strongly correlated liquid ground state. For open boundary conditions the ground state manifold consists of four degenerate sectors whose quantum numbers are determined by the orientation of the edge spins. Explicit expressions for the entanglement properties, the excitation gap as well as the exact wave functions for a couple of excited states are analytically derived and discussed.
\end{abstract}
\maketitle
In low dimensional systems strong quantum fluctuations can inhibit the formation of long range order. A particularly fascinating class where this is the case are spin liquids \cite{Balents10}. Very recent numerical theoretical work has revealed and explored spin liquid phases in two-dimensional systems with possible experimental realizability - the anti-ferromagnetic Heisenberg model on a Kagome lattice \cite{Yan11} and the frustrated XY-model on the honeycomb lattice \cite{Varney11}. An extensively studied one-dimensional system which exhibits spin liquid behavior and which is amenable to analytic treatment is the celebrated spin $1$ chain due to Affeck, Kennedy, Lieb and Tasaki (AKLT) \cite{Affleck88}. This model has no free parameters and the wave function of the ground state, which shows short-ranged entanglement and a hidden string-order, is known analytically. Initially studied in the context of valence bond solids the ground state has been demonstrated to be also of practical relevance e.g. as a resource for measurement based quantum computation \cite{Brennen08}. Moreover, the AKLT model has provided valuable guidance for spin models that cannot be treated analytically but are located in some proximity to it in the parameter space, such as the spin $1$ Heisenberg chain \cite{Miyashita93,Kolezhuk96,Polizzi98} or certain spin $1/2$ ladder systems \cite{Kim00,Kolezhuk98}. In spite of this great success and the detailed knowledge of the AKLT wave function \cite{Arovas88,Verstraete04} the analytical construction of excited state wave functions or an exact calculation of the energy gap has remained elusive.

\begin{figure}[h]
\centering
\includegraphics*[width=0.8\columnwidth]{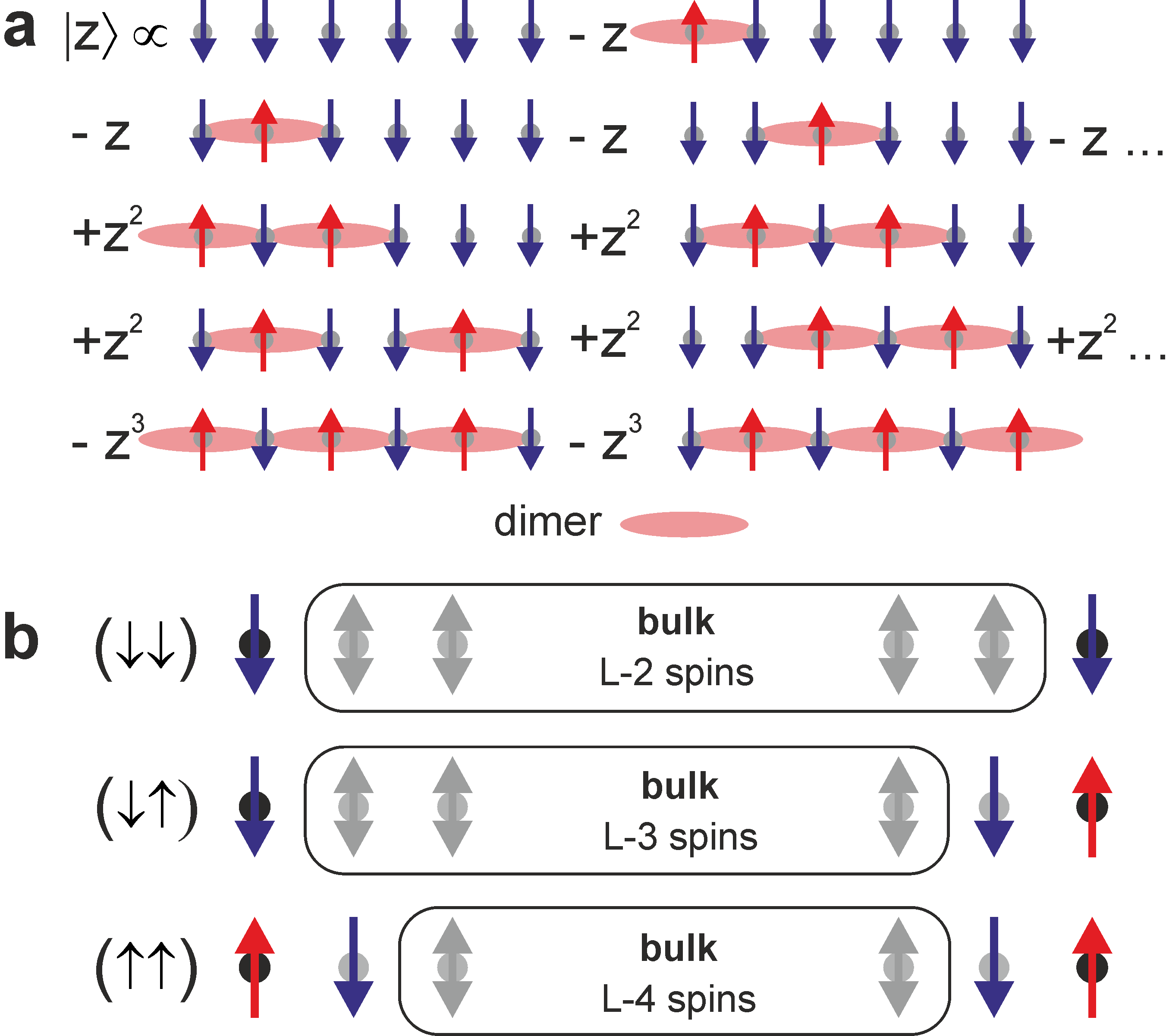}
\caption{(a) Representation of the ground state in terms of classical spin configuration (here for $L=6$ spins) which can be translated into dimer arrangements of an interstitial lattice. In general all accessible configurations contribute. The weight of each configuration is determined by the parameter $z$. (b) Chain of $L$ spins with open boundary conditions. Here the ground state is four-fold degenerate and the individual sectors are labeled by the orientation of the two edge spins $(\nu,\mu)$ with $\nu,\mu=\uparrow\downarrow$ (only three sectors are shown). Depending on the state of the edge spins the number of spins contributing to the bulk differs.}
\label{fig:edges}
\end{figure}
In this work we explore a one-parameter spin $1/2$ chain that is a limiting case of a quantum Ising model. Like the AKLT spin model it belongs to the class of frustration-free Hamiltonians \cite{deBaudrap10} and has a quantum liquid ground state with a simple matrix product state (MPS) representation \cite{Kluemper93}. We present a detailed analysis of its correlations, degeneracy and entanglement properties. More importantly, we provide analytical expressions for the energy gap and discuss the construction of the exact wave functions of a few excited states. Having access to these quantities is uncommon in non-trivial frustration-free systems with non-commuting local Hamiltonians. Here the existence of an energy gap can be usually established \cite{Spitzer03} but neither its exact value nor the structure of excited states are known. We believe that our model is particularly appealing for it is experimentally realizable, possesses a ground state with a simple MPS wave function and has a tuneable excitation gap. At the same time it is amenable to analytical treatment such that properties that go beyond the ground state can be exactly derived. It can therefore serve as a starting point for the analysis of other strongly correlated systems that depart from the exactly solvable parameter space.

\noindent\textit{Hamiltonian --- }The spin model we are considering here consists of a one-dimensional chain of $L$ spin $1/2$ particles that interact via a three-body interaction. The Hamiltonian is given by
\begin{align}
  H=H_0+H_\mathrm{b}\label{eq:H}
\end{align}
with
\begin{align}
  H_0=\sum^{L-1}_{k=2} P_{k-1}\left[\sigma^k_x+z P_k+z^{-1}n_k \right]P_{k+1}=\sum^{L-1}_{k=2} h_k\label{eq:H0}.
\end{align}
Here $P_k=(1-\sigma_z^k)/2$ is the projector on the down-state of the $k$-th spin ($\left|\downarrow\right>_k$), $\sigma^k_x$ and $\sigma^k_z$ are Pauli spin matrices and the number operator $n_k=1-P_k$ is the complement of $P_k$. Local Hamiltonians $h_k$ that belong to adjacent sites do not commute. $H_\mathrm{b}$ contains the boundary terms and reads for periodic boundary conditions $H_\mathrm{b}=P_{L}\left[\sigma^1_x+z^{-1} n_1+z P_1 \right]P_{2}+P_{L-1}\left[\sigma^L_x+z^{-1} n_{L}+z P_{L} \right]P_{1}=h_1+h_L$. In this case the Hamiltonian is symmetric under inversion and translations of the lattice sites and depends solely on the parameter $z$ which we take to be positive and real. In Ref. \cite{Lesanovsky11} it was explicitly shown how this model can be physically realized within a lattice gas of cold atoms but implementations with polar molecules \cite{Micheli06,Schachenmayer10} or trapped ions \cite{Porras04} are in principle equally possible. Those systems are governed by a Hamiltonian which maps on the quantum \emph{Ising model} in a transverse and longitudinal field with nearest and next-nearest neighbor interaction \cite{Sachdev99}:
\begin{eqnarray*}
  H_\mathrm{ph}=\sum_{k=1}^L \left[\sigma^k_x +f(z)\,n_k+ v\, n_k n_{k+1} +z\, n_k n_{k+2}\right].
\end{eqnarray*}
Hamiltonian (\ref{eq:H}) is a special case that emerges in the limit $v\gg 1$ (exclusion of neighboring excitations) and with $f(z)=z^{-1}-3z$. In the above-mentioned physical realizations the parameter $z$ can be changed through an adjustment of experimental parameters such as the strength of laser driving and the two-body interaction potential. Hamiltonian (\ref{eq:H}) possesses the $L$ conserved quantities $n_k n_{k+1}$, i.e., $\left[n_k n_{k+1},H\right]=0\,\forall\,k$. In this work we are interested in the subspace in which all of these operators have eigenvalue zero (zero-subspace), i.e. \emph{an up-spin is always accompanied by a down-spin on either side}. The previously mentioned physical implementations of $H$ are naturally confined to this subspace whose dimension is $\phi^L$ with $\phi=(1+\sqrt{5})/2$ being the \emph{golden ratio}.

\noindent\textit{Ground state --- }The ground state of Hamiltonian (\ref{eq:H}) obeys $h_k\left|z\right>=0\,\forall\,k$, i.e., it is the ground state for each of the positive semi-definite local Hamiltonians $h_k$ in eq. (\ref{eq:H}). This is the defining property of a frustration-free system \cite{deBaudrap10}. The ground state energy is hence zero and the corresponding wave function reads explicitly
\begin{eqnarray}
  \left|z\right>=\frac{1}{\sqrt{N(z)}}\prod_{k=1}^L A^\dagger_k(-z)\left|\downarrow\downarrow...\downarrow\right>.
  \label{eq:state}
\end{eqnarray}
Here the operator $A^\dagger_k(z)=\exp\left[z P_{k-1}\sigma^k_+P_{k+1}\right]$ with $\sigma_+=(\sigma_x+i\sigma_y)/2$ creates a spin in the state $\left|\downarrow\right>+z\left|\uparrow\right>$ on the $k$-th site when applied to the fully polarized state $\left|\downarrow\downarrow...\downarrow\right>$. The projection operators in the exponential of $A^\dagger_k(z)$ ensure that no two adjacent spins are simultaneously in the up-state and one thus remains in the zero subspace with respect to the operators $n_k n_{k+1}$. Furthermore, one can easily verify that $[A_k^\dagger(x),A^\dagger_m(y)]=0\,\forall\, k,m$ and $A_k^\dagger(x)A^\dagger_k(y)=A_k^\dagger(x+y)$.

The state (\ref{eq:state}) can be written as a weighted superposition of all possible dimer arrangements where an up-spin is identified as a dimer occupying an interstitial lattice as shown in Fig. \ref{fig:edges}a. The ground state is thus a superposition of a huge number of classical spin configurations as is indicated in the Figure. In the extreme case $z=1$ \emph{all spin configurations contribute with equal weight}. This is a manifestation of the strong quantum fluctuations that probe the entire accessible Hilbert space and prevent order from being formed. In the dimer picture it becomes evident that the normalization constant of the state (\ref{eq:state}) is equivalent to the partition function of a one-dimensional gas of hard dimers at fugacity $z^2$ \cite{Lesanovsky11}, and hence $N(z)=\left[(1+\sqrt{1+4z^2})/2\right]^{L}$.

Let us continue by discussing some properties of the state (\ref{eq:state}). It has a MPS representation \cite{Verstraete08} $\left|z\right>=[N(z)]^{-1/2}\sum_{i_1,..i_L=\downarrow,\uparrow}\mathrm{Tr}\left[ \, X_{i_1}X_{i_2}...X_{i_L}\right]\left|i_1,i_2,...,i_L\right>$
with the matrices $X_\uparrow=\sigma^+$ and $X_\downarrow=P-z\sigma^-$. The bond dimension of this MPS is two and entanglement is only present between nearest neighbors \cite{Wolf06}. The two spin reduced density matrix $\rho_{k k+1}$ in the basis $\left\{\left|\uparrow\right>_k\left|\downarrow\right>_{k+1},
\left|\downarrow\right>_k\left|\uparrow\right>_{k+1}, \left|\downarrow\right>_k\left|\downarrow\right>_{k+1}\right\}$ is given by
\begin{eqnarray}
  \rho_{k k+1}=\left(
              \begin{array}{ccc}
              \left<n\right> & C(z)/2 & -\left<n\right>/z \\
             C(z)/2 & \left<n\right> & -\left<n\right>/z \\
             -\left<n\right>/z & -\left<n\right>/z & 1-2\left<n\right> \\
              \end{array}
            \right).\label{eq:two_body_DM}
\end{eqnarray}
Here we have abbreviated the density $\left<n\right>=\left[1-1/\sqrt{1+4z^2}\right]/2$ and the concurrence $C(z)=2\left<n\right>\left[2\left<n\right>-1\right]/\left[\left<n\right>-1\right]$. The concurrence $0\leq C(z) \leq 1$ is directly related to the \emph{entanglement of formation} \cite{Wootters98} but also is by itself a measure of entanglement, e.g. the fully entangled singlet state $(1/\sqrt{2})[\left|\uparrow\downarrow\right>-\left|\downarrow\uparrow\right>]$ has a concurrence of one while an unentangled product state has zero concurrence. In the case of the density matrix (\ref{eq:two_body_DM}) $C(z)$ assumes its maximum value $C(z_\mathrm{max})=6-4\sqrt{2}=0.34$ at $z_\mathrm{max}=\sqrt{1+\sqrt{2}}/\sqrt{2}=1.1$.

Spins that are separated by one or more sites are not entangled and are described by product states. However, one has to distinguish here between a separation by an odd/even number of sites. The corresponding density matrices in the standard basis read $\rho^\mathrm{odd}_{ij}=[\rho_{ij}^+\otimes\rho_{ij}^++\rho_{ij}^-\otimes\rho_{ij}^-]/2$ and $\rho^\mathrm{even}_{ij}=[\rho_{ij}^+\otimes\rho_{ij}^-+\rho_{ij}^-\otimes\rho_{ij}^+]/2$
with
\begin{eqnarray}
  \rho^\pm_{ij}&=&\left(
           \begin{array}{cc}
             \left<n\right>+G_{ij} & -\left[\left<n\right>\mp G_{ij}\right]/z \\
             -\left[\left<n\right>\mp G_{ij}\right]/z & 1-\left<n\right>-G_{ij} \\
           \end{array}
         \right).
\end{eqnarray}
Here $G_{ij}=\sqrt{\left|d_{ij}\right|}$ where $d_{ij}=\left<n_i n_j\right>-\left<n_i\right>\left<n_j\right>=[z^2/(1+4z^2)][(\sqrt{1+4z^2}-2z^2-1)/(2z^2)]^{|i-j|}$ is the connected density-density correlation function. From these exponentially decaying correlations we can read off the correlation length $\xi$:
\begin{eqnarray}
  \xi^{-1}=-\log\left[1+\frac{1-\sqrt{1+4z^2}}{2z^2}\right].\label{eq:corr_length}
\end{eqnarray}
This shows that only for $z\rightarrow\infty$ long range order (without long range entanglement) is present.

\noindent\textit{Boundary conditions --- }So far all results were given for periodic boundary conditions, i.e., in the presence of $H_\mathrm{b}$, where the ground state is unique. This changes for free boundaries in which case $H_\mathrm{b}=0$. Here the Hamiltonian commutes with the operators $\sigma^1_z$ and $\sigma^L_z$, i.e. the magnetization of the edge spins is conserved. This leads to four disjoint sectors which can be labeled by the magnetization of the edge spins $(\nu,\mu)$ with $\nu,\mu=\uparrow\downarrow$ as shown in Fig. \ref{fig:edges}b. In the sector $(\downarrow,\downarrow)$ both edge spins are in the state $\left|\downarrow\right>$ and we can replace the term $h_2=P_{1}\left[\sigma^2_x+z P_2+z^{-1}n_2 \right]P_{3}$ in Hamiltonian (\ref{eq:H0}) by $\left[\sigma^2_x+z P_2+z^{-1}n_2 \right]P_{3}$ and similarly for $h_{L-1}$. The ground state of the spin chain with fixed edges is again unique, has zero energy and can be written in the form (\ref{eq:state}) with the difference that $A^\dagger_1(-z),\,A^\dagger_L(-z)\rightarrow 1$ and $A^\dagger_2(-z)\rightarrow\exp\left[-z \sigma^2_+P_{3}\right]$, $A^\dagger_{L-1}(-z)\rightarrow\exp\left[-z P_{L-2}\sigma^{L-1}_+\right]$. Hence the ground state in this sector is $\left|z\right>_{\downarrow\downarrow}=\left|\downarrow\right>_1\otimes\left|\mathrm{bulk}(z)\right>_{\downarrow\downarrow}\otimes\left|\downarrow\right>_L$ and its normalization constant is determined by the bulk wave function $N_{\downarrow\downarrow}(z)={_{\downarrow\downarrow}\!\left<\mathrm{bulk}(z)\mid\mathrm{bulk}(z)\right>_{\downarrow\downarrow}}$.
This constant is equivalent to the partition function of hard dimers placed on a line with $L-2$ sites and fugacity $z^2$. Similar arguments lead to the construction of the ground state wave functions in the remaining three sectors. However, due to $n_1n_2=n_Ln_{L-1}=0$ a spin-up state at the edge signifies that the adjacent spin has to be in the down-state which reduces the number of spins contributing to the bulk (Fig. \ref{fig:edges}b). The fact the four degenerate ground states are gapped (discussed below) and that they can be distinguished by the edge spins might make them useful for the storage of quantum information.

\noindent\textit{Spectrum --- } Let us return to periodic boundary conditions. We saw in eq. (\ref{eq:corr_length}) that the ground state exhibits density-density correlations that in general decay exponentially with the distance. Only in the limit $z\rightarrow\infty$ the correlation length reaches the entire system length. This behavior indicates the existence of an excitation gap which closes as $z$ approaches infinity.
\begin{figure}
\centering
\includegraphics*[width=1.0\columnwidth]{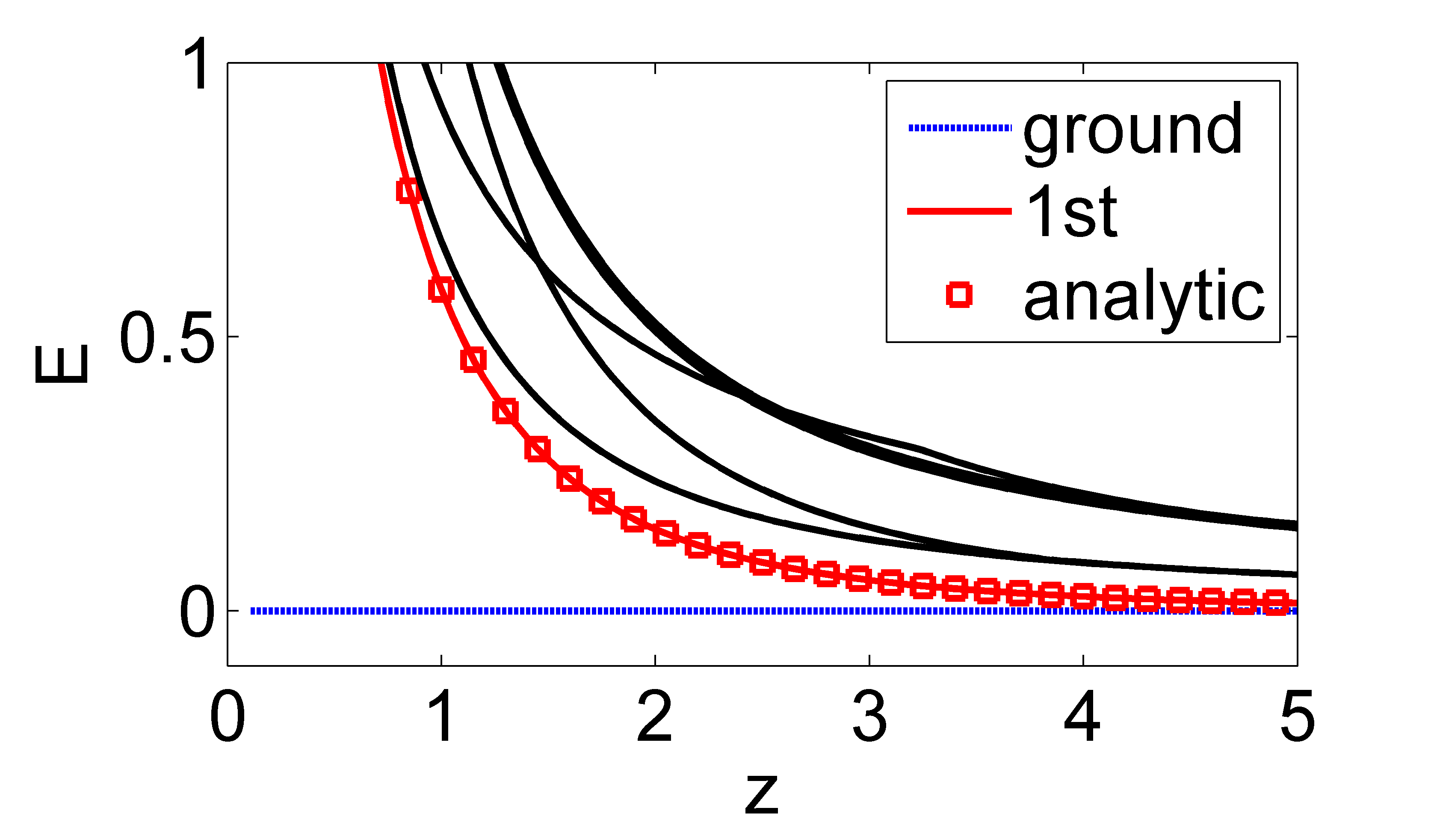}
\caption{Ground state and the first nine excited states (some of which are degenerate) of a chain of $L=16$ spins. The gap between the ground state (blue) and first excited state (red) is finite except for $z\rightarrow\infty$. The position of the red squares show the energy of the first excited state according to the analytical formula (\ref{eq:energy}).}
\label{fig:spectrum}
\end{figure}
That this is indeed the case is shown in Fig. \ref{fig:spectrum} in which displays numerical data for the energies of the ground state and the first nine excited states of a chain with 16 spins as a function of $z$. The ordered ground state at $z\rightarrow\infty$ is given by the symmetric superposition of the degenerate (anti-ferromagnetic) states $\left|u\right>=\left|\uparrow\downarrow\uparrow\downarrow...\right>$ and $\left|d\right>=\left|\downarrow\uparrow\downarrow\uparrow...\right>$.
It is now important to understand whether the gap at finite $z$ persists also in the thermodynamic limit. To get a first answer we have numerically calculated the spectrum at $z=1$ and varied the system size.
\begin{figure}
\includegraphics*[width=1.0\columnwidth]{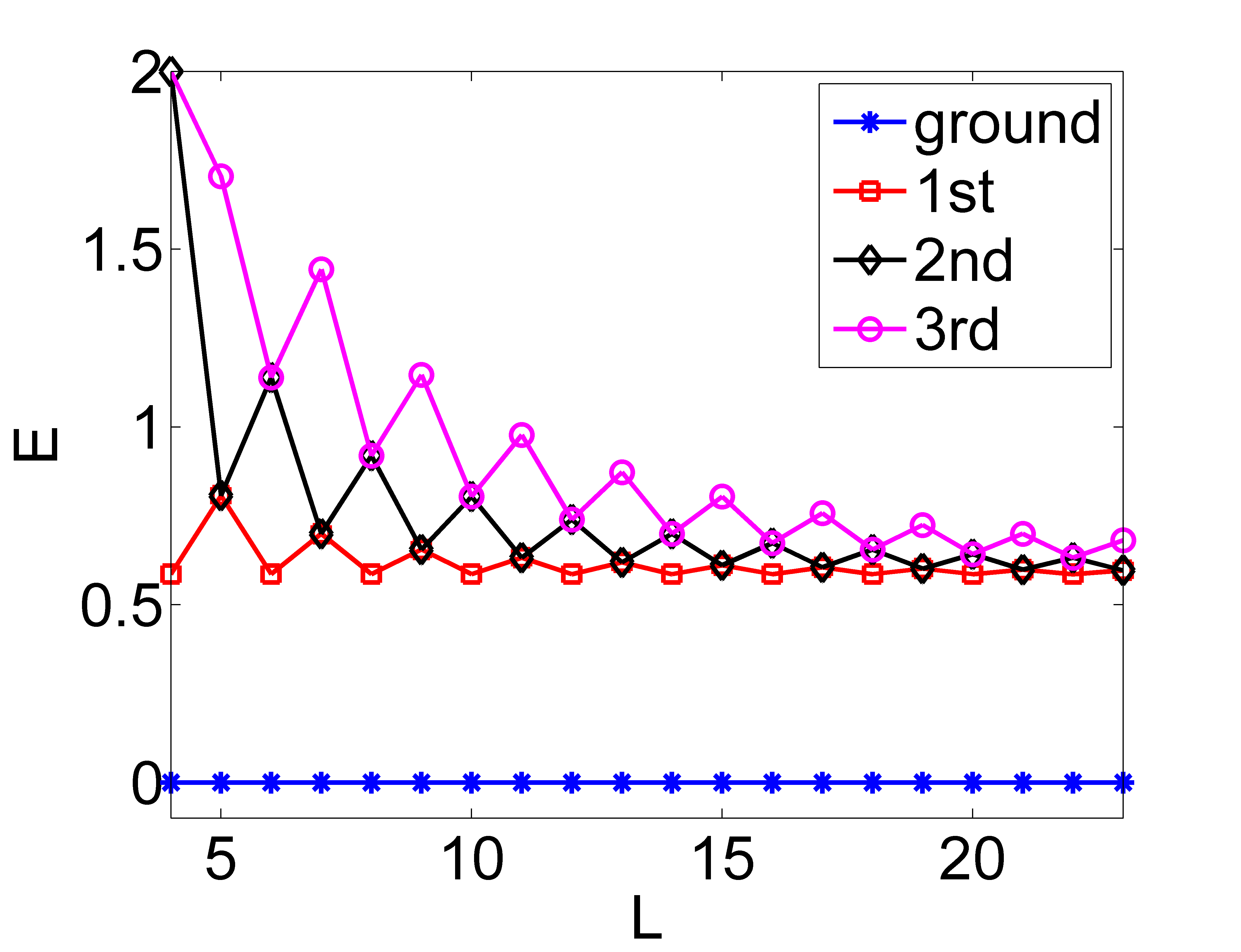}
\caption{Ground state energy and the energy of the first three excited states as a function of the system size $L$. The data is calculated for $z=1$ and periodic boundary conditions. If $L$ is even the gap is system size independent and remains constant at a value of $2-\sqrt{2}$ as predicted by eq. (\ref{eq:energy}).}
\label{fig:gap}
\end{figure}
Fig. \ref{fig:gap} shows the corresponding data. The gap persists for all values $L$ shown in the Figure while the separation between excited states decreases. This suggests that excitations are gapped for all $L$ and that the spectrum above the gap becomes continuous as $L\rightarrow\infty$. Another observation is that for even values of $L$ the excitation gap is \emph{independent of the system size}. This feature is the decisive hint for the following analytical construction of the wave function of the first excited state.

\noindent\textit{Excited states --- } We seek to construct excited states of the form $\left|E\right>=X\left|z\right>$ where $X$ is an operator that creates an excitation on the ground state. We choose the following ansatz for this operator:
\begin{eqnarray}
  X=\sum^L_{m=1}(-1)^m\left[\alpha\, n_m +\beta\, n_{m-1}n_{m+1}\right],\label{eq:ansatz}
\end{eqnarray}
where $\alpha$ and $\beta$ are real numbers. This form is motivated by the following two observations:
\\
(i) The action of off-diagonal operators on the ground state, is proportional to the action of diagonal operators: By construction we know that $h_k\left|z\right>=0$ which is the property of a frustration free Hamiltonian. Hence also $n_kh_k\left|z\right>=0$ and $P_kh_k\left|z\right>=0$. Writing out the $h_k$ explicitly and utilizing that $n_kn_{k+1}\left|z\right>=0$ one finds the following relations
\begin{eqnarray*}
  P_{k-1}\sigma^k_+P_{k+1}\left|z\right>&=&-z^{-1} n_k\left|z\right>\\P_{k-1} \sigma^k_-  P_{k+1}\left|z\right>&=&-z\, [1-n_{k-1}-n_{k}\\&&-n_{k+1}+n_{k-1}n_{k+1}]\left|z\right>
\end{eqnarray*}
This shows that an ansatz for $X$ which contains number operators and products of number operators is already the most general one.
\\
(ii) The first excited state must be antisymmetric under a translation by one lattice site: In the limit $z\rightarrow\infty$ the state (\ref{eq:state}) reduces to the symmetric superposition $\left|z\rightarrow\infty\right>=\left[\left|u\right>+\left|d\right>\right]/\sqrt{2}$,
but also all spin configurations without three adjacent down-spins, i.e. $\left|...\downarrow\downarrow\downarrow...\right>$, are equally ground states of the system as their energy tends to zero (see Fig. \ref{fig:spectrum}). The question is now which superposition of these configurations becomes the first excited state as $z$ assumes a finite value. To see this, we perform perturbation theory at large but finite $z$. Here, the term of the Hamiltonian (\ref{eq:H}) that contains $\sigma_x$ constitutes a perturbation that gives rise to virtual spin-flips, e.g.
$\left|...\downarrow\uparrow\downarrow...\right>\rightarrow\left|...\downarrow\downarrow\downarrow...\right>
\rightarrow\left|...\downarrow\uparrow\downarrow...\right>$.
A state of the ground state manifold (at $z\rightarrow\infty$) with $m$ up-spins can undergo $m$ such virtual spin flips each of which lowers the energy by $-z^{-1}$. That means the more up-spins a configuration contains, the lower its energy will be. Since this state also has to be orthogonal to $\left|z\rightarrow\infty\right>$ the antisymmetric superposition $\left[\left|u\right>-\left|d\right>\right]/\sqrt{2}$ appears to be the correct first excited state in the $z\rightarrow\infty$ limit. The antisymmetric behavior under shifts by one lattice site must be retained for all $z$ as states of different $z$ are adiabatically connected.

Using now the ansatz (\ref{eq:ansatz}) and calculating explicitly $H\left|E\right>=HX\left|z\right>=E\left|E\right>=EX\left|z\right>$ leads to a closed set of equations for the coefficients $\alpha$ and $\beta$ when $L$ is even:
\begin{eqnarray}
    \left(
     \begin{array}{cc}
       z^{-1}-z & 2z \\
       -z & 2z+2z^{-1} \\
     \end{array}
   \right)\left(
            \begin{array}{c}
              \alpha \\
              \beta \\
            \end{array}
          \right)=E\left(
            \begin{array}{c}
              \alpha \\
              \beta \\
            \end{array}
          \right).
\end{eqnarray}
The corresponding eigenvalues and eigenvectors are
\begin{eqnarray}
  E_\pm=\frac{3+z^2\pm\sqrt{1+6z^2+z^4}}{2z}\label{eq:energy}
\end{eqnarray}
and
\begin{eqnarray}
  \mathbf{v}_\pm=\left(
                   \begin{array}{c}
                     1+3z^2\mp\sqrt{1+6z^2+z^4} \\
                     2z^2 \\
                   \end{array}
                 \right)=\left(
            \begin{array}{c}
              \alpha_\pm \\
              \beta \\
            \end{array}
          \right),
\end{eqnarray}
respectively. $E_-$ corresponds to the energy of the first excited state and therefore to the excitation gap. The perfect agreement of this analytical expression with the numerical data is shown in Fig. \ref{fig:spectrum}.

Let us finally analyze the structure of the first excited state more closely. We can express its (unnormalized) wave function in terms of the operators $A^\dagger(x)$:
\begin{eqnarray}
  \left|E_-\right>&=&\sum^L_{m=1} (-1)^m A^\dagger_{m-1}\left(\gamma_-\right) A^\dagger_{m+1}\left(\gamma_-\right)\left|z\right>
\end{eqnarray}
with $\gamma_-=(2 z\beta)/(\alpha_-)$. This form suggests a pair of excitations that travel on the ground state with a lattice momentum $\pi$. To get a further idea about the appearance of this state it is also instructive to inspect the spin configurations that participate to $\left|E_-\right>$. We remember that for the liquid ground state these were in general all accessible configurations (see Fig. \ref{fig:edges}a). In contrast to this $\left|E_-\right>$ contains only configurations that differ in the number of up-spins on the two sublattices, i.e. the sublattices formed by the sites with even/odd label. Moreover, due to the second term in eq. (\ref{eq:ansatz}), configurations where the two sublattices differ also in the number of pairs of adjacent up-spins are weighted different to those where this is not the case.

\noindent\textit{Outlook --- } In this work we have analytically solved certain aspects of the model (\ref{eq:H}), but further questions remain, e.g. 'What is the nature of the (quasi-particle like) excitations?' or 'Can the model actually be solved entirely?' Concerning the latter, one can attempt to construct higher excited states by multiple applications of the operator $X$ to the ground state. However, this approach fails. It would be also interesting to see whether the treatment similar to the one presented here can be successful for other frustration-free models and/or in higher dimensions \cite{Ji11}.

We thank in E. Rico for pointing out the MPS form of the state (\ref{eq:state}) and N. Cooper, C. Ates and B. Olmos for discussions. Funding by EPSRC is acknowledged.

\end{document}